\documentclass[prl,aps,twocolumn,nofootinbib,superscriptaddress,10pt]{revtex4-1}
\usepackage{latexsym,amssymb,amsfonts,amsmath,amstext,graphicx,bbm,bm,times,relsize,dsfont,theorem,times,varioref,enumitem}
\usepackage{hyperref}
\usepackage[dvipsnames]{xcolor}
\usepackage{bm}
\usepackage{bbold}
\usepackage{array}
\usepackage{changebar}

      \colorlet{ins}{blue} \colorlet{del}{red}

\usepackage[qm]{qcircuit}
\usepackage{tikz}
\usetikzlibrary{arrows}

\newcommand{\tr}{\mbox{tr}}
\newcommand{\past}[1]{\overleftarrow{#1}}
\newcommand{\future}[1]{\overrightarrow{#1}}

\def\tr{\mbox{tr}}

\def\bra#1{\langle{#1}|}
\def\ket#1{|{#1}\rangle}
\newcommand{\braket}[2]{\left< #1 \vphantom{#2} \right|
 \left. #2 \vphantom{#1} \right>}

{\catcode`\|=\active
  \gdef\Braket#1{\begingroup
\mathcode`\|32768\let|\BraVert\left<{#1}\right>\endgroup}}
\def\BraVert{\egroup\mid\bgroup}

\newcommand{\ketbra}[2]{\ket{#1}\!\bra{#2}}

\graphicspath{{graphs/}}
\begin{document}

\title{A practical, unitary simulator for non-Markovian complex processes}

\author{Felix C. Binder}
\email{quantum@felix-binder.net}
\affiliation{School of Physical and Mathematical Sciences, Nanyang Technological University, Singapore 637371, Singapore}
\affiliation{Complexity Institute, Nanyang Technological University, Singapore 637371, Singapore}

\author{Jayne Thompson}
\affiliation{Centre for Quantum Technologies, National University of Singapore, 3 Science Drive 2, Singapore 117543, Singapore}

\author{Mile Gu}
\email{gumile@ntu.edu.sg}
\affiliation{School of Physical and Mathematical Sciences, Nanyang Technological University, Singapore 637371, Singapore}
\affiliation{Centre for Quantum Technologies, National University of Singapore, 3 Science Drive 2, Singapore 117543, Singapore}
\affiliation{Complexity Institute, Nanyang Technological University, Singapore 639673, Singapore}

\date{\today}

\begin{abstract}
Stochastic processes are as ubiquitous throughout the quantitative sciences as they are notorious for being difficult to simulate and predict. In this letter we propose a unitary quantum simulator for discrete-time stochastic processes which requires less internal memory than any classical analogue throughout the simulation. The simulator's internal memory requirements equal those of the best previous quantum models. However, in contrast to previous models it only requires a (small) finite-dimensional Hilbert space. Moreover, since the simulator operates unitarily throughout, it avoids any unnecessary information loss. We provide a stepwise construction for simulators for a large class of stochastic processes hence directly opening the possibility for experimental implementations with current platforms for quantum computation. The results are illustrated for an example process.
\end{abstract}

\maketitle

Stochastic processes permeate almost all quantitative sciences, describing the behaviour of various dynamical systems when witnessed at discrete points in time. Predictive modelling and simulation of such processes is thus of great scientific relevance, allowing inference of their future behaviour based on past observations. For complex processes, the amount of information such simulations track can become immense. 
There is thus significant interest in building simple models which can generate correct future statistics, while requiring only as much information about the past as necessary. Conceptually, such models align with the principle of Ockham's Razor, contributing to improved understanding of a process by better isolating information relevant for future prediction. Practically, all information a simulator tracks about the past must be stored, thus bounding the corresponding memory requirement for stochastic simulation.
Indeed, complexity theorists have studied this problem intensely in the field of \textit{computational mechanics}~\cite{Crutchfield1989,Shalizi2001,Shalizi2001a,Crutchfield2012}. This has resulted in a systematic method for constructing \mbox{\textit{$\varepsilon$-machines}} for general stochastic processes -- devices that replicate the future statistics of a given process while storing provably less information about past observations compared to all other alternatives. The associated amount of past information they require -- as quantified by information entropy -- is known as the \textit{statistical complexity}, a well established quantifier of structure that has been applied to diverse contexts, including neural spike trains~\cite{Haslinger2010}, geomagnetism~\cite{Clarke2003}, stock indices~\cite{Park2007,Yang2008}, crystallography~\cite{Varn2002,Varn2013,Varn2015}, and spin chains~\cite{Crutchfield1997,Suen2017,Aghamohammadi2017,Aghamohammadi2017a,Jouneghani2017}.

Quantum technologies have great potential to reshape these studies. Recent advances indicate that for almost all stochastic processes, one can construct a quantum model that demands less past information than the simplest non-quantum counterpart, while generating statistically identical future predictions~\cite{Gu2012}. This has since led to surging activity at the interface between quantum physics and computational mechanics. Notable developments include the discovery that quantum and classical statistical complexity can exhibit very different scaling and qualitative behaviour~\cite{Suen2017,Aghamohammadi2017,Jouneghani2017}, a proposal of more efficient quantum models whose memory saving scale with long-range correlations present in a stochastic process~\cite{Mahoney2016,Riechers2016}, and a recent application to sampling of rare events~\cite{Aghamohammadi2018}. Each of these reflects growing evidence that what we consider complex can fundamentally change in the quantum regime.

To fully establish such observations, however, one must also develop an understanding of the mechanisms in which quantum models for stochastic processes generate future predictions. For instance, in addition to a quantum model's statistical complexity, it is interesting to also consider the number and dimension of physical systems dedicated to its working memory as well as the type of required interactions between these systems. 
Meanwhile, to achieve optimal results the currently simplest known quantum models require measurement in a Hilbert space whose dimension scales with the \textit{cryptic order} of a stochastic process -- a property closely related to the presence of long-range correlations~\cite{Mahoney2016}. This space would become impractically large with increasing cryptic order.

This letter aims to address these issues. We propose a quantum simulator for stochastic processes that includes an efficient quantum circuit for the generation of future predictions. It satisfies three particular desiderata. (1) The amount $C_q$ of past information required by the simulator equals that of the most efficient quantum models known so far. (2) The circuit is unitary, such that the internal entropy of the simulator remains $C_q$ throughout. (3) Future predictions are generated using only bi-partite interactions between two systems of bounded dimension (assuming bounded $C_q$ and finite output alphabet).

Taken together, the presented results close a loophole whereby reduced entropic memory may not necessarily lead to practical savings of memory dimension in stochastic simulation. By showing that the simulator's memory cost never exceeds $C_q$ they complement the work by Riechers et al. who developed a method for computing $C_q$~\cite{Riechers2016}.
Lastly, they directly open the door for practical simulation of more complex stochastic processes in quantum laboratories.\pagebreak

\begin{figure}[ht]
 \centering
 \includegraphics[width=0.46\textwidth]{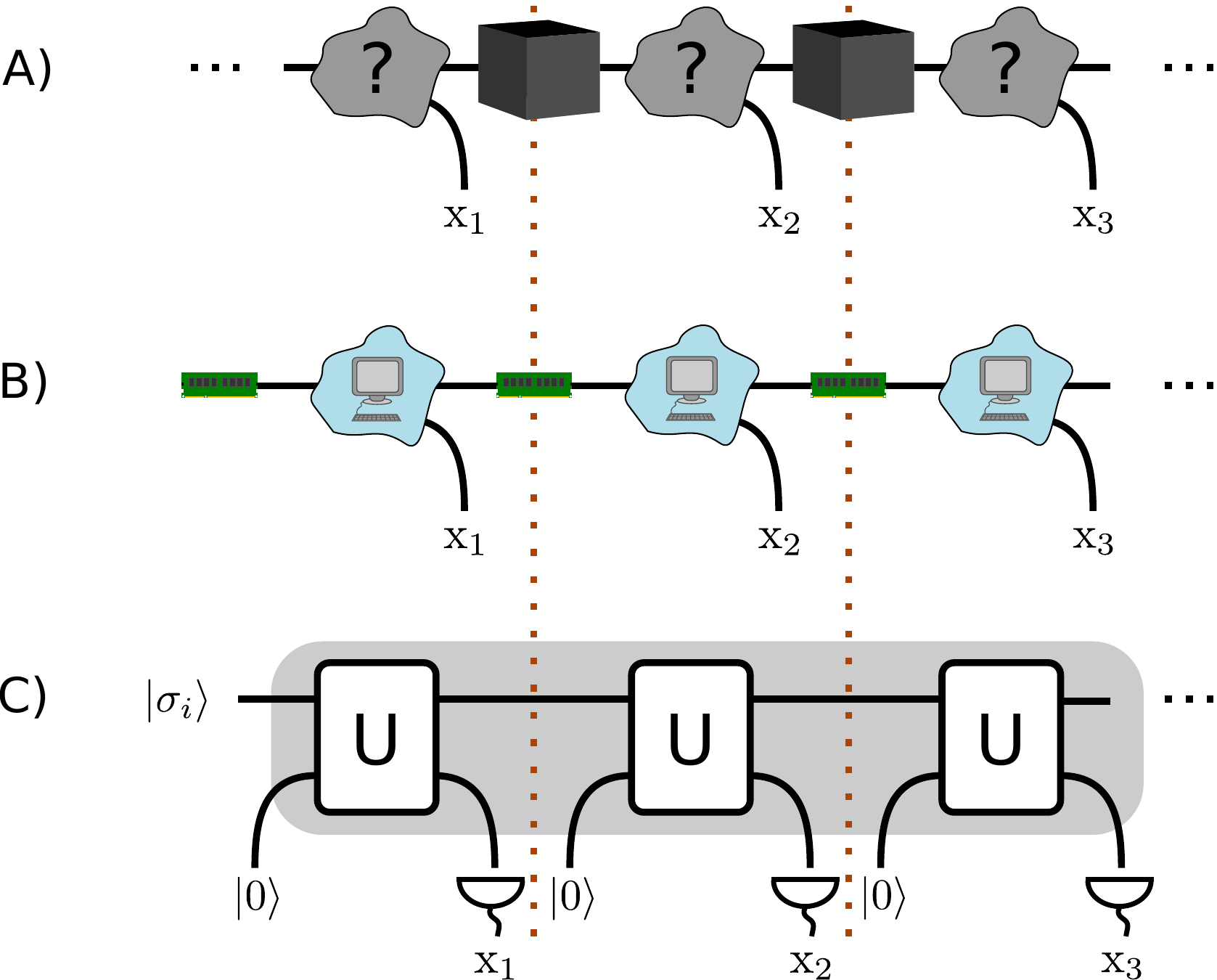}
 \caption{A stochastic process (A), a general simulator (B), and a q-simulator (C) generally all retain some memory about the past (brown, dotted lines) between each symbol emission.\\
 A) A stochastic process is a bi-infinite string of random variables, $\ldots, X_{-2}, X_{-1}, X_0, X_{1}, X_{2}, \ldots$. We picture it here as `emitting' symbols $x_t$ at discrete times $t$.
 \\
 B) The process' simulator equally emits symbols $x_t$ at times $t$. Its output is statistically indistinguishable from the process itself.
 \\
 C) A q-simulator is a specific type of simulator which uses quantum memory, initialised in state $\ket{\sigma_i}$ with probability $\pi_i$ (top register), and a sequentially-repeated unitary interaction with one or multiple symbol registers. It may be thought of as a 'black box' (represented in grey) which takes states $\ket{0}$ as input and returns output states whose measurement statistics in the computational basis are indistinguishable from the process itself. The physical system which supplies the blank $\ket{0}$-states can be reused at each step.
 }
  \label{fig:simulator}
\end{figure}

\noindent\textbf{Computational mechanics.}
A stochastic process is a bi-infinite sequence of random variables $X_t$, labelled by discrete time-steps $t$~\cite{Gallager2013} (see Fig.~\ref{fig:simulator}A). It is governed by a probability distribution $P(\past{X},\future{X})$ where $\past{X} \equiv \ldots, X_{-2},X_{-1},X_0$ corresponds to the \textit{past} of a process and $\future{X} \equiv X_{1},X_{2},\ldots$ its \textit{future}. Each $X_t$ takes values $x$ from an \textit{alphabet} $\mathcal{A}$. Strings of output symbols $x_{m:n}:=x_{m+1}x_{m+2}..x_n$ form \textit{words}. Each instance of a process has a specific past $\overleftarrow{x}\equiv x_{-\infty:0}$, with each specific future $\overrightarrow{x}\equiv x_{0:\infty}$ occurring according to the conditional probability $P(\future{X} = \overrightarrow{x}|\overleftarrow{x})$. Processes are assumed to be \textit{stationary}, such that they are invariant with respect to time-translations~\footnote{Formally, stationarity implies that for any word $w_L$ consisting of $L$ symbols $P(X_{0:L}=w_L)=P(X_{t:t+L}=w_L)\;\forall\;t, L$.\\This does not contradict the fact that generally $P(X_{0:L}=w_L|x_{-\infty:0})\neq P(X_{t:t+L}=w_L|x_{-\infty:t})$ for $x_{-\infty:0}\neq x_{-\infty:t}$.
}. We hence omit the corresponding index $t$.

A simulator replicates the conditional behaviour of the corresponding process (Fig.~\ref{fig:simulator}B). It is represented by a hidden Markov model (HMM)~\cite{Rabiner1986}: For each past $\overleftarrow{x}$, the simulator's memory can be configured in some internal state $s_{\overleftarrow{x}}$, such that its future outputs are statistically indistinguishable from the stochastic process that is being modelled. Once configured in $s_{\overleftarrow{x}}$, the simulator sequentially emits symbols governed by the expected conditional probability distribution $P(\overrightarrow{X}|\overleftarrow{x})$. To do so it needs to update the state of its memory at each time-step. For instance, on emitting a symbol $r$ a simulator in state $s_{\overleftarrow{x}}$ would transition to state $s_{\overleftarrow{x}'}$ corresponding to the updated past $\overleftarrow{x}' = \overleftarrow{x} r$.

Computational mechanics singles out a specific type of HMM called \textit{$\epsilon$-machine} by application of Ockham's razor -- i.e., the internal states should contain as much information about the past as is required for accurate simulation of the future but nothing more~\cite{Crutchfield1989,Shalizi2001,Shalizi2001a,Crutchfield2012}. 
This is formalised by a mapping $\varepsilon\!:\mathcal{A}^\infty\to\Sigma$ which partitions the set of pasts into a set of internal states $\Sigma=\{s_1, s_2, \ldots\}$ according to the following equivalence relation:
\begin{equation}
 \overleftarrow{x}\sim_\varepsilon \overleftarrow{x}'\text{, iff }P(\overrightarrow{X}|\overleftarrow{x})=P(\overrightarrow{X}|\overleftarrow{x}').
 \label{eq:equivalence}
\end{equation}

Accordingly, all pasts which imply identical future statistics correspond to the same internal state (called \textit{causal state} in this context).

For illustration, consider the \textit{alternating process} -- a string of alternating symbols $1$ and $0$. Its future may be replicated by storing only the last symbol. This process hence has two causal states, in correspondence with the last observed value.

Eq.~\ref{eq:equivalence} entails a property called \textit{unifilarity}. It implies that the current causal state, together with the next output symbol, determines the subsequent causal state with certainty, i.e., by some function $\lambda(i,x)$ which takes the value of the next causal state's label. Hence, the conditional transition probabilities $P(x,j|i)$ -- corresponding to a transition from state $s_i$ to state $s_j$ upon emission of symbol $x$ -- can be expressed in terms of $P(x|i)$, the probability of emitting symbol $x$ from causal state $i$:
\begin{equation}
 P(x,j|i)=P(x|i)\delta_{j,\lambda(i,x)}\label{eq:Kronecker},
\end{equation}
where $\delta_{j,k}$ is the Kronecker-$\delta$.

In practice, an $\varepsilon$-machine's defining quantities $\mathcal{A}$, $\Sigma$, $P(x|i)$, and $\lambda(i,x)$ may be obtained from a given data string by a suitable reconstruction algorithm~\cite{Crutchfield1989,Shalizi2002,Strelioff2014}.

It follows from stationarity that $P(x|i)$ and $\lambda(i,x)$ do not change over time. Moreover, the \textit{stationary probability} $\pi_j$ of being in causal state $s_j$ is time-independent.

This allows the definition of the \textit{statistical complexity}
\begin{equation}
 C_\mu:=-\sum_i\pi_i\log \pi_i\label{eq:cmu}.
\end{equation}
It represents the average amount of information an $\varepsilon$-machine needs to store about the past~\cite{Crutchfield1989,Grassberger1986,Shalizi2001}. Alternatively, the \textit{topological complexity}
\begin{equation}
 C_\mu^0:=\log|\Sigma|\label{eq:c0}
\end{equation}
quantifies the memory of an $\varepsilon$-machine in terms of the number of causal states $|\Sigma|$~\cite{Crutchfield1989}.

\noindent\textbf{Q-simulators and their internal states.} We now describe q-simulators -- fully-quantum simulators for stationary stochastic processes that extend $\varepsilon$-machines to the quantum domain. Illustrated in Fig.~\ref{fig:simulator}C, they are a specific type of simulator comprising a single quantum system as internal memory and functioning by repeated application of the same unitary operation $U$.

A q-simulator for a given stochastic process features a set of internal states $\{\ket{\sigma_i}\}$, each associated with the process' corresponding causal state. These states may be used to unitarily generate a future pattern of any length, one symbol at a time (Fig.~\ref{fig:simulator}C).

Starting in $\ket{\sigma_i}$, the total state of the circuit after the first application of $U$ is given by
\begin{equation}
 \ket{1_i}:=U\ket{\sigma_i}\ket{0} =\sum_x\sqrt{P(x|i)}\ket{\sigma_{\lambda(i,x)}}\ket{x}, \label{eq:1}
\end{equation}
where the sum goes over all symbols $x$. Using Eq.~\ref{eq:Kronecker} we may equally write $\ket{1_i}=\sum_x\sum_j\sqrt{P(x,j|i)}\ket{\sigma_j}\ket{x}$.

More generally, after $L$ applications, the total state corresponding to the first $L+1$ registers in Fig.~\ref{fig:simulator}C is
\begin{equation}
\label{eq:L}
\begin{split}
 \ket{L_i}&:=U^L\ket{\sigma_i}\ket{0}^{\otimes L}\\
 &=\sum_{x_{0:L}}\sqrt{P({x_{0:L}}|i)}\ket{\sigma_{\lambda(i,{x_{0:L}})}}\ket{x_{0:L}}.
\end{split}
\end{equation}
Here, $\lambda(i,x_{0:L})$ extends previous notation in a natural way from single symbols to words. $U$ only acts on two registers at a time (see Fig.~\ref{fig:simulator}C); identity operations on the remaining registers are omitted.

When the symbol-registers are measured in the computational basis (i.e., given by $\{\ket{x}\}$) each word $x_{0:L}$, corresponding to a single state $\ket{\sigma_{\lambda(i,{x_{0:L}})}}$ in the sum, is measured with the desired probability $P(x_{0:L}|i)$. This extends to mixtures of different $\ket{L_i}$.

It remains to be shown that the desired operator $U$ actually exists. Considering states $\ket{L_i}, \ket{L_j}, \ket{M_i},$ and $\ket{M_j}$, as defined by Eq.~\ref{eq:L}, this is the case iff $\braket{L_i}{L_j}=\braket{M_i}{M_j}$ for all $L,M,i,j$, or -- more simply -- iff
\begin{equation}
 \braket{\sigma_i}{\sigma_j}=\braket{1_i}{1_j}\;\forall\;i,j.
 \label{eq:inner-products}
\end{equation}
For any unitary operator $U$ which fulfils $U\ket{\sigma_j}\ket{0}=\ket{1_j}\;\forall\; j$, Eq.~\ref{eq:inner-products} is also fulfilled as a consequence of unitarity. Conversely, if Eq.~\ref{eq:inner-products} is not fulfilled there does not exist a unitary $U$ as required by Eqns.~\ref{eq:1} and~\ref{eq:L}~\cite{Jozsa2000}. In the following section, we describe a constructive procedure for obtaining $U$ which starts from Eq.~\ref{eq:inner-products}.

First, however, we consider the memory required for such unitary simulation, i.e., in the top register of Fig.~\ref{fig:simulator}C, initialised in state $\ket{\sigma_i}$ with probability $\pi_i$ (See Supplementary Information~A). As in the classical case, this memory may be quantified in two ways: by its entropic information content (\textit{quantum machine complexity} $C_q$) and by the logarithm of the required memory Hilbert space dimension ($C_q^0$) which always upper-bounds $C_q$\footnote{Operationally, $C_q$ may be understood as the asymptotic memory requirement for storing the current causal states of many parallel, independent and identically distributed processes. For $N\to\infty$ processes, the number of required qubits scales as $NC_q$~\cite{Schumacher1995}.\\$C_q^0$, on the other hand, quantifies the memory required in a single run in terms of the memory Hilbert space dimension.}. Extending Eqns.~\ref{eq:cmu},~\ref{eq:c0} to the quantum case:

\begin{align}
  C_q:=&S\left[\sum_j\pi_j\ket{\sigma_j}\!\bra{\sigma_j}\right],\label{eq:cq}\\ 
  C_q^0:=&\log\left(\text{rank}\left[\sum_j\pi_j\ket{\sigma_j}\!\bra{\sigma_j}\right]\right). \label{eq:cq0}
\end{align}

In general, $C_q\leq C_\mu$, $C_q^0\leq C_\mu^0$, and the ratios $C_\mu/C_q$ and $C_\mu^0/C_q^0$ can become arbitrarily large (see Supplementary Information~B for an example of $C_\mu/C_q\to\infty$). Moreover, there are processes for which $C_q(\zeta)$ (or $C_q^0(\zeta)$) attains a finite asymptotic value while $\lim C_\mu(\zeta)\to\infty$ (or $\lim C_\mu^0(\zeta)\to\infty$) as a function of some system parameter $\zeta$~\cite{Aghamohammadi2017a,Elliott2018,Garner2017,Thompson2017a}.

While the present focus lies on $C_q$ -- in line with previous literature -- the results equally apply to $C_q^0$. Importantly, there exist processes with $C_q^0< C_\mu^0$ and even unbounded advantage has been shown for $C_q^0$~\cite{Thompson2017a}. Such processes can be unitarily simulated with a memory Hilbert space of dimension $\exp[C_q^0]$, according to the method presented in the following section.

In Supplementary Information~C it is proven that Eq.~\ref{eq:cq} coincides with the memory requirements of the simplest previously known quantum models~\cite{Mahoney2016,Riechers2016}.

\noindent\textbf{Unitary generation.} The unitary operator associated with each q-simulator is obtained as follows, starting from the representation of a given $\varepsilon$-machine in terms of $\mathcal{A}$, $\Sigma$, $P(x|i)$, and $\lambda(i,x)$:
\begin{enumerate}[label=(\Roman*)]
 \item Express $\ket{1_i}$ in terms of $\ket{\sigma_j}$ (Eqns.~\ref{eq:1}) and solve Eqns.~\ref{eq:inner-products} for the inner products $c_{ij}\equiv\braket{\sigma_i}{\sigma_j}$.
 \item On a Hilbert space of dimension $|\Sigma|$ choose some basis $\{\ket{e_j}\}$ and express $\{\ket{\sigma_j}\}$ in an inverse Gram-Schmidt procedure, i.e.~\footnote{For some processes, the internal states $\ket{\sigma_j}$ may be linearly-dependent. This does not impede the generality of our construction: We may set $U$ to be the identity operation on that subspace which is not spanned by $\{\ket{\sigma_j}$\}.}~\footnote{A similar construction was presented in Supplementary Information~C of~\cite{Riechers2016}. In that case, however, subsequent memory states are not generally unitarily connected.}:
\begin{align}
 \ket{\sigma_1}=\ket{e_1};\label{eq:GS}\; \ket{\sigma_2}&=c_{12}\ket{e_1}+\sqrt{1-c_{12}^2}\ket{e_2};
 \; \text{etc.}
\end{align}
 \item Solve Eq.~\ref{eq:1} for those columns of $U$ which correspond to input states $\ket{e_j}\ket{0}$. This determines the matrix elements $U_{ij,i'0}\equiv\bra{e_i}\bra{j}U\ket{e_{i'}}\ket{0}$.
 \item Obtain the remaining columns of $U$ by a Gram-Schmidt procedure starting from the already-determined columns.
\end{enumerate}

This prescription is constructive, and works for all discrete-valued stationary processes with a finite number of causal states (see Supplementary Information~D).

\noindent\textbf{Example: upset-gambler process.}\label{page:example} We demonstrate our construction explicitly for the upset-gambler process. The process is distinguished as being particularly simple but yet having infinitely long memory (more technically: cryptic order, cf. Eq.~\ref{eq:k}). Its $\varepsilon$-machine consists of two causal states, $s_A$ and $s_B$. The non-zero transition probabilities are, in shorthand, $P(0,s_B|s_A)=p$, $P(1,s_A|s_A)=1-p$, $P(0,s_A|s_B)=q$, and $P(1,s_A|s_B)=1-q$. It is assumed that $0<p<1$, $0<q<1$, and $p< q$. Further details are included in Supplementary Information~B. 

Replacing Arabic with Roman numerals for the states of the upset-gambler process we have:
\begin{align}
 \ket{I_A}&=\sqrt{1-p}\ket{\sigma_A}\ket{1}+\sqrt{p}\ket{\sigma_B}\ket{0}\text{, and}\\
 \ket{I_B}&=\sqrt{q}\ket{\sigma_A}\ket{0}+\sqrt{1-q}\ket{\sigma_A}\ket{1}.
\end{align}
Eqns.~\ref{eq:inner-products},~\ref{eq:GS} 
allow the internal states to be expressed in terms of any orthonormal qubit basis $\{\ket{e_j}\}$:
\begin{align}
 \ket{\sigma_A}=\ket{e_1};\;
 \ket{\sigma_B}=c\sqrt{\bar{p}\bar{q}}\ket{e_1}+c \xi\ket{e_2}\label{eq:sigmas}
\end{align}
with $\bar p\equiv1-p$, $\bar q\equiv 1-q$, $c\equiv(1-\sqrt{pq})^{-1}$, $\xi\equiv\sqrt{q}-\sqrt{p}$. This in turn allows for the specification of the first and third column of the desired unitary matrix $U$, written in the basis $\{\ket{e_1}\ket{0},\ket{e_1}\ket{1},\ket{e_2}\ket{0},\ket{e_2}\ket{1}\}$:
\begin{equation}
U= \left(\begin{array}{cccc}
c\sqrt{p\bar p \bar q} & \sharp  & 1- cp\bar q & \sharp\\
\sqrt{\bar p} & \sharp & -\sqrt{p\bar q} & \sharp\\
c\xi\sqrt{p} & \sharp & -c\sqrt{p\bar p \bar q} & \sharp\\
0 & \sharp & 0 & \sharp
\end{array}\right).
\end{equation}
Here, $\sharp$ stands for undetermined matrix entries. The only condition for determining these entries is that the four columns of the matrix must form an orthonormal basis. For instance, a simple choice would be $(0,0,0,1)^T$ for the second column and $(\sqrt{p\bar q}, -\sqrt{pq}, \sqrt{\bar p},0)^T$ for the fourth.

Using this 2-qubit unitary operator, the upset-gambler process can be simulated with internal entropy $C_q$ throughout. The latter may be easily computed using Eqns.~\ref{eq:cq},~\ref{eq:sigmas}, and the stationary probabilities $\pi_A=1/(1+p)$, $\pi_B=p/(1+p)$.

\noindent\textbf{Relation to previous works.} Q-simulators coincide with previous models for the case of Markovian processes~\cite{Gu2012, Mahoney2016, Riechers2016} where unitary implementation has already been demonstrated~\cite{Palsson2017,Jouneghani2017}. For non-Markovian processes, however, the currently simplest known quantum models face a trade-off between memory reduction due to quantum encoding and increasing size of the simulator's working memory (proportional to the number $L$ of $|\mathcal{A}|$-dimensional systems  constituting said memory)~\cite{Mahoney2016,Riechers2016}. This trade-off can be seen in those models' quantum machine complexity:
\begin{align}
   \tilde{C}_q(L)&:=S\left[\sum_i\pi_i\ket{\eta_i(L)}\!\bra{\eta_i(L)}\right]\label{eq:cql}\text{, where}\\
 \ket{\eta_i(L)}&:=\sum_{x_{0:L}}\sqrt{P(x_{0:L}|i)}\ket{\lambda(i,x_{0:L})}\ket{x_{0:L}}.\label{eq:eta}
\end{align}
Here, orthonormal states $\ket{j}$ (i.e., $\ket{\lambda(i,x_{0:L})}$) correspond to classical causal states $s_j$.

While this has not been proven in general, there is strong numerical indication that $\tilde{C}_q(L)$ decreases with increasing encoding length $L$~\cite{Riechers2016}. No further reduction occurs when $L$ exceeds the cryptic order, formally defined as the smallest $k$ for which the following conditional Shannon entropy vanishes:
\begin{equation}
H(S_{k}|\overrightarrow{X})=0.\label{eq:k}
\end{equation}
Here, $S_k$ is the random variable corresponding to the causal state after $k$ time-steps. Hence, $\tilde{C}_q(k)=\tilde{C}_q(M)\;\forall\;M>k$. In the case of infinite cryptic order, $\tilde{C}_q(L)$ approaches $\lim_{L\to\infty}\tilde{C}_q(L)$ exponentially~\cite{Riechers2016,Mahoney2016}.

\begin{figure}[ht]
 \includegraphics[width=\linewidth]{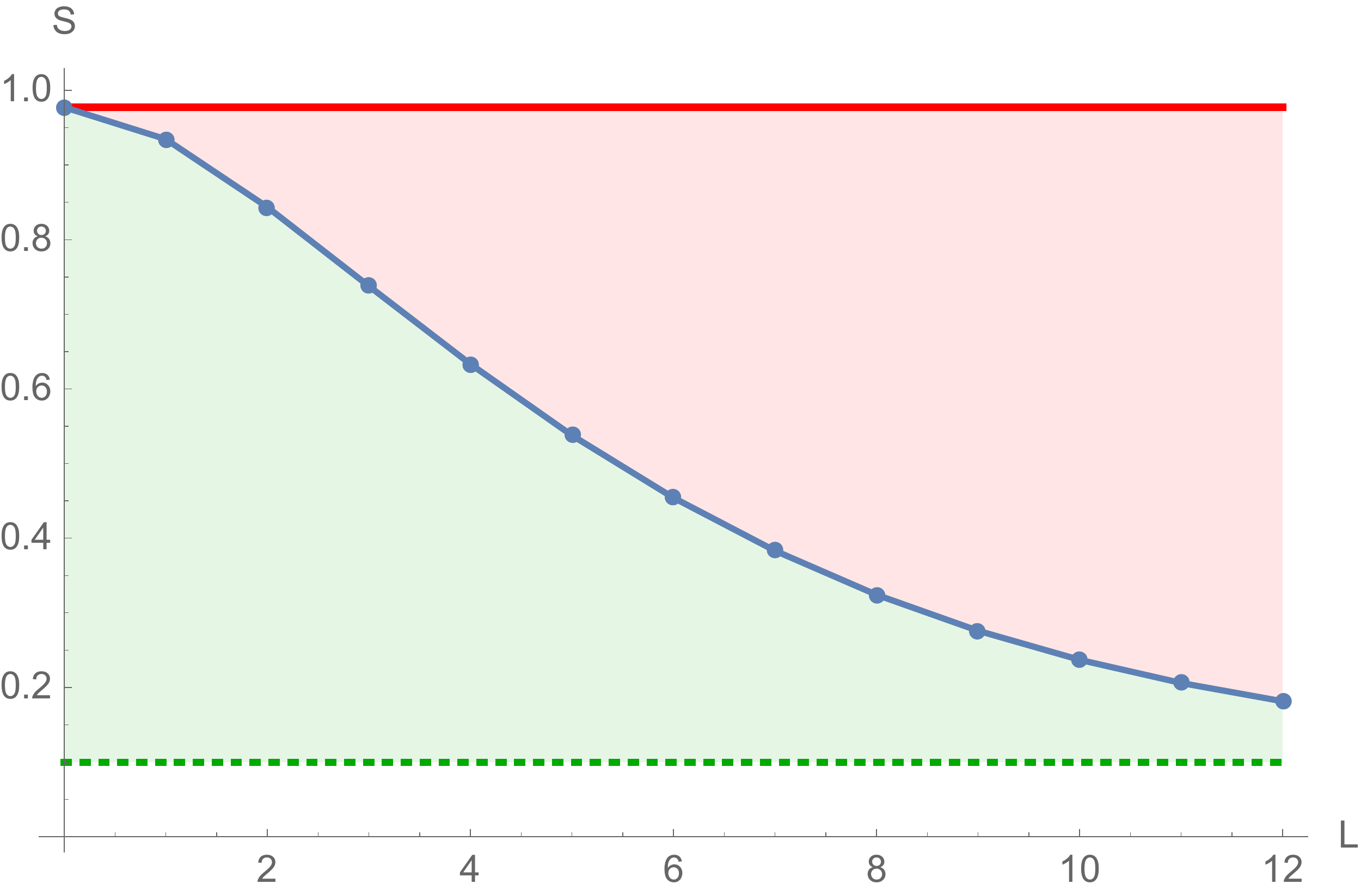}
 \caption{Memory requirement $C_q$ for simulation of the upset-gambler process with $p=0.7$ and $q=0.8$. The solid red line shows the classical complexity $C_\mu\approx0.98$. The blue dots represent the memory $\tilde C_q(L)$ required by previous models. $L$ corresponds to the number of future symbols stored at a time and, hence, equivalently, the number of qubits required for the simulator's internal memory (see Eqns.~\ref{eq:cql},~\ref{eq:eta}). $\tilde C_q(L)$ is always lower bounded by $C_q\approx 0.10$ which it reaches asymptotically for $L\to\infty$. The q-simulator operates at $C_q$ throughout (green, dashed).\\For an illustration of the process' full parameter range see Supplementary Information~B.
 }
 \label{fig:L-word-Cq}
\end{figure}

Supplementary Information~C contains the proof that $C_q=\lim_{L\to\infty}\tilde{C}_q(L)$. Importantly, q-simulators only require a single system of dimension $|\mathcal{A}|$ as working memory to achieve the same memory savings. This is shown in Fig.~\ref{fig:L-word-Cq} for the example of the upset-gambler process.

In addition to a trade-off between memory savings and dimension of the internal working memory, previous models for non-Markovian processes implicitly rely on a measure-and-prepare logic~\cite{Gu2012,Mahoney2016,Riechers2016,Aghamohammadi2018}. They encode causal states in distinct orthogonal states of the simulator's internal memory (i.e., the states $\ket{j}$ in Eq.~\ref{eq:eta}) and require projective measurement and a fresh encoding for the next simulation step. For instance, if causal state $\ket{j}$ were measured, the state $\ket{\eta_j(L)}$ would be prepared. Between measurement and preparation the memory required to store the outcome consequently equals the classical statistical complexity $C_\mu$. Q-simulators add to these models by allowing for unitary simulation even of non-Markovian processes, hence confirming that $C_q =\lim_{L\to\infty}\tilde{C}_q(L)$ correctly identifies the amount of quantum memory throughout the simulation and not just at intermediate points. Since q-simulators operate by sequential application of the same unitary interaction at each time-step, they require no additional processing of the memory system.

\noindent\textbf{Discussion.} We have introduced unitary quantum simulators of stochastic processes which store less information than any classical counterpart. The resulting q-simulator encodes relevant information about a process' past directly into states retained in quantum memory. At each time-step $t$, the q-simulator unitarily interacts its memory and an incoming ancillary system, such that later measurement of that system in a standard basis yields a statistically correct output $x_t$, and simultaneously collapses the memory into the correct quantum state for continued simulation at the next time-step. Our framework is constructive: Given a stochastic process, one can solve for both the quantum memory-states and the exact form of the unitary gate allowing future simulation. The resulting quantum simulator's required memory, $C_q$, aligns with the best previously known quantum constructions~\cite{Mahoney2016,Riechers2016,Gu2012}. Our new simulation scheme then improves upon these by providing a means to generate desired predictions, one step at a time, by repeated unitary interaction such that the memory of the simulator never exceeds $C_q$ at any time during the operation. It hence avoids trade-offs between entropic memory reduction and dimension of the memory state space. Complementing reference~\cite{Riechers2016}, our results may be used to analytically compute $C_q$. The same construction holds when the single-shot memory quantifier $C_q^0$ is considered, further reducing the memory Hilbert space for processes which exhibit $C_q^0<C_\mu^0$.

These results pave the way to new experimental realisations. Quantum advantage in stochastic simulation has been demonstrated for the special case of Markovian processes~\cite{Palsson2017,Jouneghani2017}, but dimensional scaling has so far made simulations of non-Markovian counterparts significantly more challenging. Application of our results to an infinite Markov order process has resulted in a q-simulator that requires only a single two-qubit unitary gate and a single qubit of memory, bringing it well within reach of present experimental technologies.

It would be interesting to extend these results to  continuous- time processes~\cite{Marzen2016,Elliott2018}, and to interactive systems which take information from their environment and use it to generate suitable output responses~\cite{Barnett2015,Thompson2017}. In both cases quantum models have been demonstrated to be more powerful than all classical counterparts. Our techniques offer a promising approach towards building unitary quantum simulators for such generalised scenarios.

These further research avenues will benefit from recent results that relate stochastic processes to so-called matrix product states (MPS) which are used in the description of quantum spin chains~\cite{Yang2018}.

\noindent\textbf{Acknowledgements.} The authors thank A. Monras, K. Wiesner, C. Yang, V. Narasimhachar, C. Di Franco, A. Garner, T. Elliott, W. Y. Suen, N. Tischler, J. Mahoney, and J. Crutchfield for valuable comments and acknowledge support by the National Research Foundation of Singapore (Fellowship NRF-NRFF2016-02), the John Templeton Foundation (Grant No. 54914), the Foundational Questions Institute (grant ``Observer-dependent complexity: the quantum-classical divergence over `what is complex?'''), and the Singapore Ministry of Education (Tier 1 grant RG190/17).

\clearpage

\renewcommand{\thepage}{\roman{page}}
\setcounter{page}{1}
\setcounter{equation}{0}
\setcounter{figure}{0}

        \renewcommand{\theequation}{S\arabic{equation}}%
        \renewcommand{\thefigure}{S\arabic{figure}}%
\begin{widetext}

\section*{Supplementary Information~A: memory requirement}
A q-simulator's memory register is initialised in state $\ket{\sigma_i}$ with the corresponding stationary probability $\pi_i$. The eigenvalues of the resulting mixed state
\begin{equation}
 \phi:=\sum_i \pi_i\ketbra{\sigma_i}{\sigma_i}
\end{equation}
define the quantum statistical complexity $C_q$ (Eq.~\ref{eq:cq}) and the quantum topological complexity $C_q^0$ (Eq.~\ref{eq:cq0}). More generally, any Renyi entropy $H^\alpha(\phi)$ may equally be calculated in terms of $\phi$'s eigenvalues.

In this section we prove that $\phi$ is a fixed point of the channel $\Lambda$ which $U$ induces on the memory register:
\begin{equation}
 \Lambda(\rho):=\text{tr}_S\left[U\rho\otimes\ketbra{0}{0}U^\dag\right].
\end{equation}
Here, $\tr_S$ denotes a partial trace over the symbol register.
\begin{equation}
\begin{split}
 \Lambda(\phi)&=\tr_S\left[U\left(\sum_i\pi_i\ketbra{\sigma_i}{\sigma_i}\right)\otimes\ketbra{0}{0}U^\dag\right]\\
	      &=\sum_i\pi_i\;\tr_S\ketbra{I_i}{I_i}\\
	      &=\sum_i\pi_i\sum_y\sum_{x,x',j,j'}\sqrt{P(x,j|i)P(x',j'|i)}\braket{y}{x}\ketbra{\sigma_j}{\sigma_{j'}}\braket{x'}{y}\\
	      &=\sum_i\pi_i\sum_x\sum_{j,j'}\sqrt{P(x,j|i)P(x,j'|i)}\ketbra{\sigma_j}{\sigma_{j'}}\\
	      &=\sum_i\pi_i\sum_{x,j}P(x,j|i)\ketbra{\sigma_j}{\sigma_j}\\
	      &=\sum_j\underbrace{\left(\sum_x\sum_i\pi_i P(x,j|i)\right)}_{=P(j)}\ketbra{\sigma_j}{\sigma_j}\\
	      &=\sum_j\pi_j\ketbra{\sigma_j}{\sigma_j}\\
	      &=\phi
\end{split}
\end{equation}
In the first three lines, we have used Eqns.~\ref{eq:Kronecker}~and~\ref{eq:1}. The fifth line follows from the fourth by unifilarity (see Eq.~\ref{eq:Kronecker}) and the last three lines follow, first, from basic probability theory (using $P(a)=\sum_bP(a|b)P(b)=\sum_bP(a,b)$) and then from stationarity of the process (i.e., $\pi_j=P(j)$ in the present shorthand notation). \hfill$\square$

This result implies that the memory requirement of a q-simulator may be characterised by any Renyi-entropy $H^\alpha(\phi)$ when the simulator operates on the stationary state. The quantum statistical complexity $C_q$ ($\alpha\to 1$) and the quantum topological complexity $C_q^0$ ($\alpha=0$) are of particular physical relevance.

If the register corresponding to the output symbol is measured, the total entropy of memory and symbol register decreases (equally for multiple measured output registers). Importantly, this implies that the entropy of the memory register -- initialised in $\phi$ -- is bounded from above by $C_q$ when conditioned on measurement results on the output registers.

\clearpage


\section*{Supplementary Information~B: the upset-gambler process}
The upset-gambler process is depicted in Fig.~\ref{fig:upset-gambler-2}.  It is assumed that $0<p<1$, $0<q<1$. The restriction to $q>p$, which was introduced for simplicity of notation in the main text, is here replaced by the more general assumption $p\neq q$. Note that for $p=q$ this process would reduce to a biased coin, comprising only a single causal state. The process statistics remain non-Markovian as $p$ approaches $q$. However, in this limit the upset-gambler statistics become more and more similar to a biased coin.
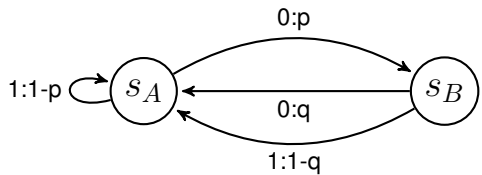
\begin{figure}[ht]
\centering
\begin{tikzpicture}[->,>=stealth',shorten >=1.5pt,auto,node distance=4cm,thick,main node/.style={circle,draw,font=\sffamily\Large\bfseries}]
\centering

 \node[main node] (1) {$s_A$};
 \node[main node] (2) [right of=1] {$s_B$};

 \path[every node/.style={font=\sffamily}]
 (1) edge [bend left] node {0:p} (2)
     edge [loop left] node {1:1-p} (1)
 (2) edge [bend left] node {1:1-q} (1)
     edge node {0:q} (1);
 \end{tikzpicture}
 \caption{$\varepsilon$-machine representation of the upset-gambler process. This process has infinite cryptic order. Each edge is labelled by the emitted symbol followed by its probability of emission.}
 \label{fig:upset-gambler-2}
\end{figure}

This process describes a skilled gambler who loses with probability $p$ (output: $0$). When she loses she becomes upset and plays one game with a different losing probability $q$. Figs.~\ref{fig:upset-gambler-3D} and \ref{fig:upset-gambler-2D} show $C_\mu$ and $C_q$ for the full parameter range of the upset-gambler process ($0<p<1$, $0<q<1$, $p\neq q$).
\begin{figure}[ht]
 \includegraphics[width=0.73\linewidth]{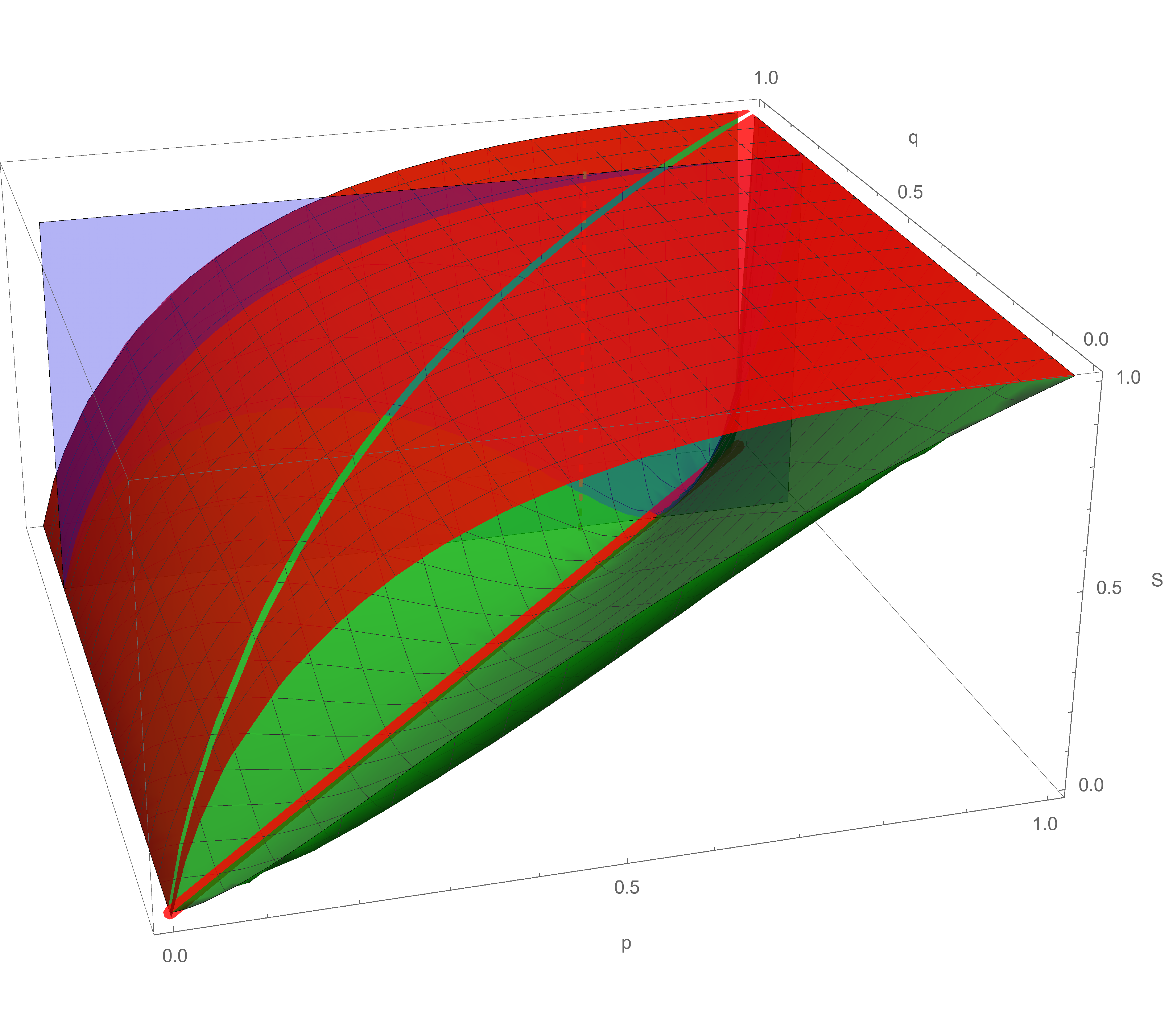}
 \caption{$C_\mu$ (red) and $C_q$ (green) as functions of the parameters $p$ and $q$. For $p=q$ the process becomes a biased coin which has no statistical complexity. This value is reached discontinuously in the case of $C_\mu$ and continuously for $C_q$. The brown line indicates the parameter values which were used in Fig.~\ref{fig:L-word-Cq}. The blue intersecting plane is shown in Fig.~\ref{fig:upset-gambler-2D}.}
 \label{fig:upset-gambler-3D}
\end{figure}

Despite its simplicity, the upset-gambler process exhibits some remarkable properties which are symptomatic for a large group of complex processes:
\begin{enumerate}
 \item It has infinite cryptic and Markov order. That is, the memory time of the process is infinitely long. This is indicated by the fact that the same words of any length can be emitted starting from either one of the causal states, $s_A$ or $s_B$.
 \item It shows what has been called the ``ambiguity of simplicity'' \cite{Suen2017,Aghamohammadi2017,Jouneghani2017}. This is clearly visible in Fig.~\ref{fig:upset-gambler-2D} where $C_q$ has negative slope between its local maximum at $p\approx 0.28$ and $p\to0.8$ while $C_\mu$ has positive slope in the same parameter range. For instance, $C_q(p=0.4)>C_q(p=0.6)$ but $C_\mu(p=0.4)<C_\mu(p=0.6)$.\\This entails that the (statistical) complexity one should attribute to a stochastic process depends on the type of memory substrate (quantum or classical).
 \item For $p\to q$ the ratio $C_\mu/C_q$ -- termed \textit{quantum advantage} in~\cite{Aghamohammadi2018}-- diverges, implying unbounded memory savings when using quantum rather than classical memory to simulate a large number of independent, identical upset-gambler processes with $p$ and $q$ approaching equality. This may be clearly observed in Fig.~\ref{fig:upset-gambler-2D} (see blue, dotted curve).
\end{enumerate}

\begin{figure}[hb]
 \includegraphics[width=0.73\linewidth]{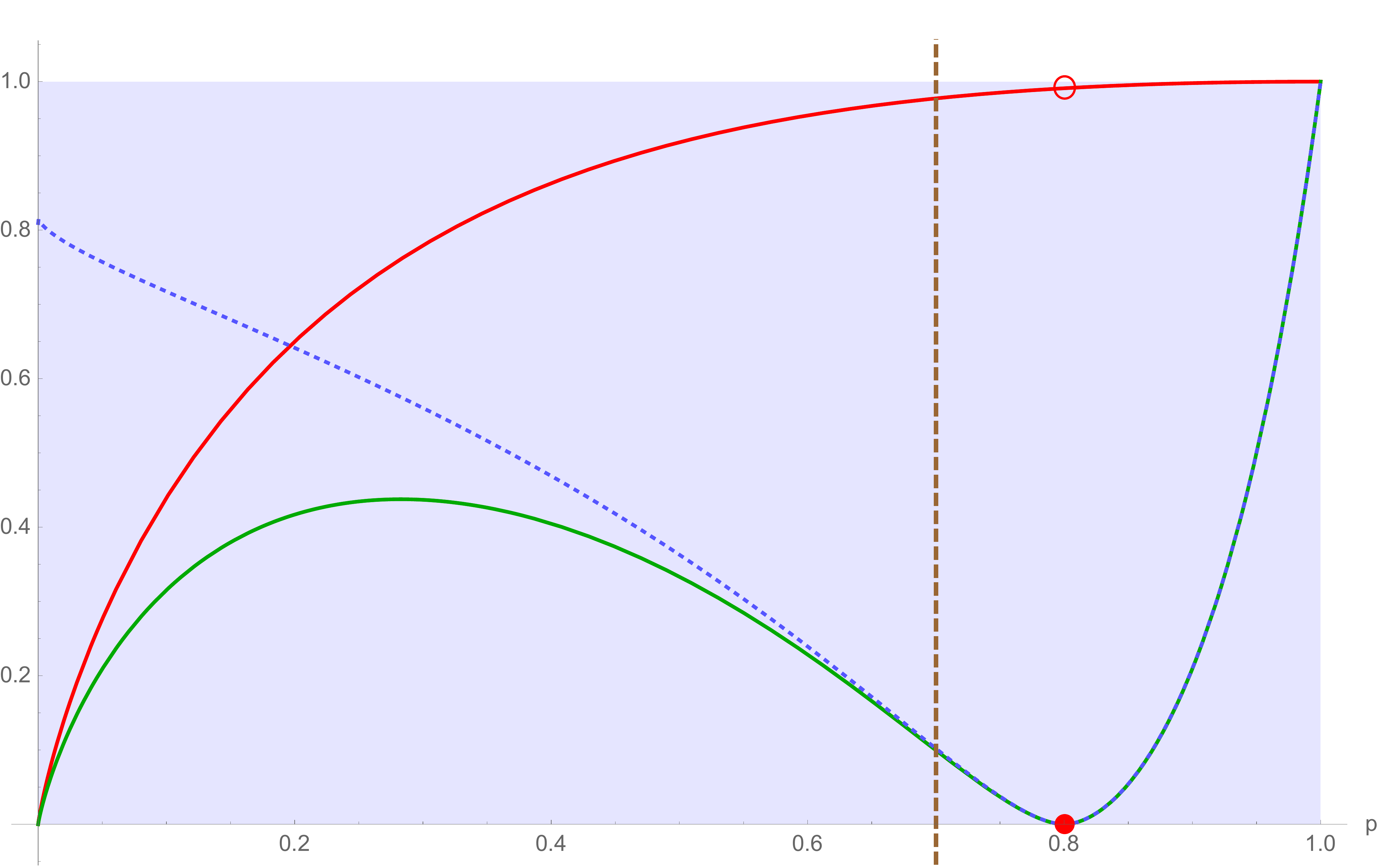}
 \caption{This figure corresponds to the blue plane in Fig.~\ref{fig:upset-gambler-2D} and shows $C_\mu$ (red) and $C_q$ (green) as functions of the parameter $p$, with $q=0.8$. $C_\mu$ is discontinuous at $p=q$. The blue, dotted curve represents the ratio $C_q/C_\mu$ whose inverse diverges for $p\to 0.8$. The parameter values which were used in Fig.~\ref{fig:L-word-Cq} are indicated by the brown, dashed line.}
 \label{fig:upset-gambler-2D}
\end{figure}


\clearpage


\section*{Supplementary Information~C: $\tilde C_q(L)$ converges towards $C_q$}
The memory required by a q-simulator may be quantified, as is commonly done, by the quantum statistical complexity $C_q$ (Eq.~\ref{eq:cq}) for the assymptotic case of many independent and identical processes being simulated in parallel. In a single run the quantum topological complexity $C_q^0$ (Eq.~\ref{eq:cq0}) may serve as an alternative quantifier of the simulator's memory (see Supplementary Information~A).

$C_q$ is straightforward to compute with the approach described in the main text. In this section we show that it also corresponds to the smallest quantum statistical complexity previously reported in other works~\cite{Mahoney2016,Riechers2016}:
\begin{align}
    C_q&=\lim_{L\to\infty}\tilde{C}_q(L)\text{, where}\label{eq:C_equality}\\
   \tilde{C}_q(L)&:=S\left[\sum_j\pi_j\ket{\eta_j(L)}\!\bra{\eta_j(L)}\right]\tag{\ref{eq:cql}}\text{, with}\\
 \ket{\eta_i(L)}&:=\sum_{x_{0:L}}\sqrt{P(x_{0:L}|i)}\ket{\lambda(i,x_{0:L})}\ket{x_{0:L}}.\tag{\ref{eq:eta}}
\end{align}

\noindent Consider Eq.~\ref{eq:L} again, which is our model's analogue to Eq.~\ref{eq:eta}:
\begin{equation}
 \begin{split}
  \ket{L_i}:=&U^L\ket{\sigma_i}\ket{0}^{\otimes L}\\
  =&\sum_{x_{0:L}}\sqrt{P(x_{0:L}|i)}\ket{\sigma_{\lambda(i,x_{0:L})}}\ket{x_{0:L}},
 \end{split}\tag{\ref{eq:L}}
\end{equation}
Note that in the first line of this equation $U$ only acts on the space of internal states and that of one symbol at a time. As before, identity operations on the remaining symbol spaces have been omitted for clarity.
Since $\braket{L_i}{L_j}=\braket{\sigma_i}{\sigma_j}\;\forall\;i,j,L$ the numerical value of Eq.~\ref{eq:cq} does not change if the states $\ket{\sigma_j}$ are replaced with the corresponding states $\ket{L_j}$. Hence, in order to prove Eq.~\ref{eq:C_equality} it is in fact sufficient to show that
\begin{equation}
 \lim_{L\to\infty}(\underbrace{\braket{L_i}{L_j}-\braket{\eta_i(L)}{\eta_j(L)}}_{\equiv \delta(i,j,L)})=0\;\forall\;i,j.
\end{equation}
The difference inside this limit can be bounded as follows (Each step is explained in the paragraphs below the equation):
\begin{equation}
\begin{split}
 \delta(i,j,L)=&\sum_{x_{0:L}}\sqrt{P(x_{0:L}|i)P(x_{0:L}|j)}\left[\braket{\sigma_{\lambda(i,x_{0:L})}}{\sigma_{\lambda(j,x_{0:L})}}-\delta_{\lambda(i,x_{0:L}),\lambda(j,x_{0:L})}\right]\\
 =&\sum_{\{x_{0:L}:\lambda(i,x_{0:L})\neq\lambda(j,x_{0:L})\}}\sqrt{P(x_{0:L}|i)P(x_{0:L}|j)}\left[\braket{\sigma_{\lambda(i,x_{0:L})}}{\sigma_{\lambda(j,x_{0:L})}}-\delta_{\lambda(i,x_{0:L}),\lambda(j,x_{0:L})}\right]\\
 \leq&\sum_{\{x_{0:L}:\lambda(i,x_{0:L})\neq\lambda(j,x_{0:L})\}}\sqrt{P(x_{0:L}|i)P(x_{0:L}|j)}\\
 =&\sum_{i'\neq j'}\bra{(s_i,s_j)}\zeta^L\ket{(s_{i'},s_{j'})}.\label{eq:overlapDiff}
\end{split}
\end{equation}
From the first to the second line the sum was restricted to only those words $x_{0:L}$ which lead to different causal states when emitted from the initial states $s_i$ and $s_j$, respectively. In all cases where a word leads to the same state the difference inside the brackets vanishes. For finite Markov order $R\leq L$ this happens for all $i$, $j$, and $x_{0:L}$ (Markov order is defined as the smallest $R$ for which the following conditional Shannon entropy $H$ vanishes: $H[S_R|X_{0:R}]=0$). The proof is hence already complete for cases of finite Markov order. Its remainder is hence only concerned with processes of infinite Markov order.

The third line follows from the second (1) by virtue of the fact that the $\delta$-function vanishes for all elements left in the sum, and (2) due to the first term in the bracket being non-negative and upper-bounded by unity for all $i\neq j$ (Equality holds only if the sum vanishes at finite $L$).

The step from the third to the fourth line requires some further definitions. First, the matrix $\zeta$ acts on the space of causal state pairs $(s_i,s_j)$ represented by orthonormal states $\ket{(s_i,s_j)}$. It is very similar to the identically-labelled matrix which appears in reference \cite{Riechers2016} in the context of the \textit{quantum pairwise merger machine (QPMM)}. We follow a similar line of argument here (with the difference that the present $\zeta$ is a $|\Sigma|^2\times|\Sigma|^2$-matrix and there is no 'sink' state, where $|\Sigma|$ is the number of causal states). The matrix elements are
\begin{equation}
 \zeta_{(i,j),(i',j')}:=\sum_{x}\sqrt{P(x|i)P(x|j)}\delta_{i',\lambda(i,x)}\delta_{j',\lambda(j,x)}.
\end{equation}
Here, each pair of indices $(i,j)$ is understood as a row index of the matrix and analogously, $(i',j')$ is a column index. It is easily confirmed that this implies the following entries of the matrix $\zeta^L$:
\begin{equation}
 \left(\zeta^L\right)_{(i,j),(i',j')}:=\sum_{x_{0:L}}\sqrt{P(x_{0:L}|i)P(x_{0:L}|j)}\delta_{i',\lambda(i,x_{0:L})}\delta_{j',\lambda(j,x_{0:L})}.
\end{equation}
\noindent First, note that since all elements of $\zeta^L$ are positive it is certainly column-substochastic for all $\varepsilon$-machines with infinite Markov order and $L>|\Sigma|$:
\begin{equation}
\sum_{(i',j')}\left(\zeta^L\right)_{(i,j),(i',j')}<1\;\;\;\forall\;(i,j)
\end{equation}
This can be seen as follows:
\begin{equation}
 \begin{split}
 \sum_{(i',j')}\left(\zeta^L\right)_{(i,j),(i',j')}=&\sum_{(i',j')}\sum_{x_{0:L}}\sqrt{P(x_{0:L}|i)P(x_{0:L}|j)}\delta_{i',\lambda(i,x_{0:L})}\delta_{\lambda(j,x_{0:L}),j'}\\
  =&\sum_{x_{0:L}}\sqrt{P(x_{0:L}|i)P(x_{0:L}|j)}\\
  \leq&\sum_{x_{0:L}}P(x_{0:L}|i)P(x_{0:L}|j)\\
  \leq&\sum_{x_{0:L}}P(x_{0:L}|i)\sum_{x'_{0:L}}P(x'_{0:L}|j)\\
  =&1
 \end{split}
 \label{eq:substochasticity}
\end{equation}
Equality between the first and the last line holds if and only if the corresponding $\varepsilon$-machine involves deterministic causal state transition chains from causal states $s_i$ and $s_j$ under emission of the same word of length $L$ ($s_i$ may be the same as $s_j$). However, it is easily confirmed that as long as $L>|\Sigma|$ any such $\varepsilon$-machine must have Markov order $R=|\Sigma|$.

Hence, for all $\varepsilon$-machines of finite $|\Sigma|$ and infinite Markov order, $\zeta^L$ is row-substochastic for $L>|\Sigma|$. Gerschgorin's theorem applies to such matrices and it directly implies that all singular values, and hence also the spectral radius are strictly smaller than unity~\cite{Gerschgorin1931}.

This in turn implies
\begin{equation}
 \lim_{L\to\infty}\zeta^L=0.
\end{equation}
Hence, Eq.~\ref{eq:overlapDiff} vanishes for infinite Markov order as well as for finite Markov order (as already proved above). This proves Eq.~\ref{eq:C_equality} for all cases where the number of causal states is finite: A q-simulator's entropic memory requirement $C_q$ is as small as that of the best previous predictive models, i.e., $\lim_{L\to\infty}\tilde{C}_q(L)$. \hfill$\square$

In reference~\cite{Riechers2016} it was shown that all pairwise overlaps $\braket{\eta_i(L)}{\eta_j(L)}$ monotonically decrease with increasing $L$. It was pointed out in the same reference that this does not mathematically imply that $\tilde C_q(N)\leq\tilde C_q(M)$ for $N>M$. While strong evidence of such monotonic convergence was given, a general proof remains, as of yet, elusive.

\clearpage


\section*{Supplementary Information~D: Existence of solutions for algorithm to construct $\{\ket{\sigma_i}\}$ and $U$}
In the main text we gave an algorithm for the construction of the set of internal memory states $\ket{\sigma_i}$ and the unitary operator $U$. For the sake of completeness, in this section we show that step I) always has a solution and comment on what happens in step II) when the states $\ket{\sigma_i}$ are linerarly dependent. Steps III) and IV) of the algorithm require no further explanation.

\begin{enumerate}[label=(\Roman*)]
\item We show that a solution exists by specifying it explicitly. Introducing the notation $\overrightarrow {x_t}:=x_{t+1}x_{t+2}\ldots$ for infinite-length words, a solution to Eq.~\ref{eq:inner-products} is given by the following assignment:
\begin{equation}
 c_{ij}=\sum_{\overrightarrow{x_0}} \sqrt{P(\overrightarrow {x_0}|i)P(\overrightarrow {x_0}|j)}.
 \label{eq:cij}
\end{equation}
where $P(\overrightarrow{x_t}|i)$ denotes the probability of emitting the infinite-length word $\overrightarrow{x_t}$ when starting in causal state $s_i$. Note that $c_{ij}=\sum_{\overrightarrow{x_t}} \sqrt{P(\overrightarrow {x_t}|i')P(\overrightarrow {x_t}|j')}\;\forall\;t$, due to stationarity of the underlying process. We express Eq.~\ref{eq:inner-products} explicitly and insert Eq.~\ref{eq:cij} to show self-consistency:
\begin{equation}
\begin{split}
 c_{ij}\equiv\braket{\sigma_i}{\sigma_j}&=\sum_{x,i',j'}\sqrt{P(x,i'|i)P(x,j'|j)}c_{i'j'}\\
 &=\sum_{i',j'}\sum_x\sqrt{P(x,i'|i)P(x,j'|j)}\sum_{\overrightarrow{x_1}} \sqrt{P(\overrightarrow {x_1}|i')P(\overrightarrow {x_1}|j')}\\
 &=\sum_{\overrightarrow{x_0}}\underbrace{\sum_{i'}\sqrt{P(x,i'|i)}\sqrt{P(\overrightarrow{x_1}|i')}}_{=\sqrt{P(\overrightarrow{x_0}|i)}}\underbrace{\sum_{j'}\sqrt{P(x,j'|j)}\sqrt{P(\overrightarrow{x_1}|j')}}_{=\sqrt{P(\overrightarrow{x_0}|j)}}\\
 &=\sum_{\overrightarrow{x_0}} \sqrt{P(\overrightarrow {x_0}|i)P(\overrightarrow {x_0}|j)}\\
 &=c_{ij}
 \end{split}
\end{equation}

\item The construction is self-explanatory but it is worth highlighting that the states $\ket{\sigma_i}$ are not necessarily linearly-independent. In this case only $r$ of the basis states $\ket{e_i}$ will appear in the construction, where $r<|\Sigma|$ is the rank of $\{\ket{\sigma_i}\}$. In Eq.~\ref{eq:GS}
, the prefactors of states $\ket{e_s}$ with $s>r$ all simply vanish and the general construction is hence equally valid in this case.
\end{enumerate}
\clearpage
\end{widetext}


\begin{thebibliography}{36}%
\makeatletter
\providecommand \@ifxundefined [1]{%
 \@ifx{#1\undefined}
}%
\providecommand \@ifnum [1]{%
 \ifnum #1\expandafter \@firstoftwo
 \else \expandafter \@secondoftwo
 \fi
}%
\providecommand \@ifx [1]{%
 \ifx #1\expandafter \@firstoftwo
 \else \expandafter \@secondoftwo
 \fi
}%
\providecommand \natexlab [1]{#1}%
\providecommand \enquote  [1]{``#1''}%
\providecommand \bibnamefont  [1]{#1}%
\providecommand \bibfnamefont [1]{#1}%
\providecommand \citenamefont [1]{#1}%
\providecommand \href@noop [0]{\@secondoftwo}%
\providecommand \href [0]{\begingroup \@sanitize@url \@href}%
\providecommand \@href[1]{\@@startlink{#1}\@@href}%
\providecommand \@@href[1]{\endgroup#1\@@endlink}%
\providecommand \@sanitize@url [0]{\catcode `\\12\catcode `\$12\catcode
  `\&12\catcode `\#12\catcode `\^12\catcode `\_12\catcode `\%12\relax}%
\providecommand \@@startlink[1]{}%
\providecommand \@@endlink[0]{}%
\providecommand \url  [0]{\begingroup\@sanitize@url \@url }%
\providecommand \@url [1]{\endgroup\@href {#1}{\urlprefix }}%
\providecommand \urlprefix  [0]{URL }%
\providecommand \Eprint [0]{\href }%
\providecommand \doibase [0]{http://dx.doi.org/}%
\providecommand \selectlanguage [0]{\@gobble}%
\providecommand \bibinfo  [0]{\@secondoftwo}%
\providecommand \bibfield  [0]{\@secondoftwo}%
\providecommand \translation [1]{[#1]}%
\providecommand \BibitemOpen [0]{}%
\providecommand \bibitemStop [0]{}%
\providecommand \bibitemNoStop [0]{.\EOS\space}%
\providecommand \EOS [0]{\spacefactor3000\relax}%
\providecommand \BibitemShut  [1]{\csname bibitem#1\endcsname}%
\let\auto@bib@innerbib\@empty
\bibitem [{\citenamefont {Crutchfield}\ and\ \citenamefont
  {Young}(1989)}]{Crutchfield1989}%
  \BibitemOpen
  \bibfield  {author} {\bibinfo {author} {\bibfnamefont {J.~P.}\ \bibnamefont
  {Crutchfield}}\ and\ \bibinfo {author} {\bibfnamefont {K.}~\bibnamefont
  {Young}},\ }\href {\doibase 10.1103/PhysRevLett.63.105} {\bibfield  {journal}
  {\bibinfo  {journal} {Physical Review Letters}\ }\textbf {\bibinfo {volume}
  {63}},\ \bibinfo {pages} {105} (\bibinfo {year} {1989})}\BibitemShut
  {NoStop}%
\bibitem [{\citenamefont {Shalizi}\ and\ \citenamefont
  {Crutchfield}(2001)}]{Shalizi2001}%
  \BibitemOpen
  \bibfield  {author} {\bibinfo {author} {\bibfnamefont {C.~R.}\ \bibnamefont
  {Shalizi}}\ and\ \bibinfo {author} {\bibfnamefont {J.~P.}\ \bibnamefont
  {Crutchfield}},\ }\href {\doibase 10.1023/A:1010388907793} {\bibfield
  {journal} {\bibinfo  {journal} {Journal of Statistical Physics}\ }\textbf
  {\bibinfo {volume} {104}},\ \bibinfo {pages} {817} (\bibinfo {year}
  {2001})},\ \Eprint {http://arxiv.org/abs/9907176} {arXiv:9907176 [cond-mat]}
  \BibitemShut {NoStop}%
\bibitem [{\citenamefont {Shalizi}(2001)}]{Shalizi2001a}%
  \BibitemOpen
  \bibfield  {author} {\bibinfo {author} {\bibfnamefont {C.~R.}\ \bibnamefont
  {Shalizi}},\ }\emph {\bibinfo {title} {{Causal Architecture, Complexity and
  Self-Organization in Time Series and Cellular Automata}}},\ \href {\doibase
  citeulike-article-id:1028937} {Ph.D. thesis},\ \bibinfo  {school} {University
  of Wisconsin-Madison} (\bibinfo {year} {2001})\BibitemShut {NoStop}%
\bibitem [{\citenamefont {Crutchfield}(2012)}]{Crutchfield2012}%
  \BibitemOpen
  \bibfield  {author} {\bibinfo {author} {\bibfnamefont {J.~P.}\ \bibnamefont
  {Crutchfield}},\ }\href {\doibase 10.1038/nphys2190} {\bibfield  {journal}
  {\bibinfo  {journal} {Nature Physics}\ }\textbf {\bibinfo {volume} {8}},\
  \bibinfo {pages} {17} (\bibinfo {year} {2012})}\BibitemShut {NoStop}%
\bibitem [{\citenamefont {Haslinger}\ \emph {et~al.}(2010)\citenamefont
  {Haslinger}, \citenamefont {Klinkner},\ and\ \citenamefont
  {Shalizi}}]{Haslinger2010}%
  \BibitemOpen
  \bibfield  {author} {\bibinfo {author} {\bibfnamefont {R.}~\bibnamefont
  {Haslinger}}, \bibinfo {author} {\bibfnamefont {K.~L.}\ \bibnamefont
  {Klinkner}}, \ and\ \bibinfo {author} {\bibfnamefont {C.~R.}\ \bibnamefont
  {Shalizi}},\ }\href {\doibase 10.1162/neco.2009.12-07-678} {\bibfield
  {journal} {\bibinfo  {journal} {Neural Computation}\ }\textbf {\bibinfo
  {volume} {22}},\ \bibinfo {pages} {121} (\bibinfo {year} {2010})},\ \Eprint
  {http://arxiv.org/abs/1001.0036} {arXiv:1001.0036} \BibitemShut {NoStop}%
\bibitem [{\citenamefont {Clarke}\ \emph {et~al.}(2003)\citenamefont {Clarke},
  \citenamefont {Freeman},\ and\ \citenamefont {Watkins}}]{Clarke2003}%
  \BibitemOpen
  \bibfield  {author} {\bibinfo {author} {\bibfnamefont {R.~W.}\ \bibnamefont
  {Clarke}}, \bibinfo {author} {\bibfnamefont {M.~P.}\ \bibnamefont {Freeman}},
  \ and\ \bibinfo {author} {\bibfnamefont {N.~W.}\ \bibnamefont {Watkins}},\
  }\href {\doibase 10.1103/PhysRevE.67.016203} {\bibfield  {journal} {\bibinfo
  {journal} {Physical Review E}\ }\textbf {\bibinfo {volume} {67}},\ \bibinfo
  {pages} {016203} (\bibinfo {year} {2003})},\ \Eprint
  {http://arxiv.org/abs/0110228} {arXiv:0110228 [cond-mat]} \BibitemShut
  {NoStop}%
\bibitem [{\citenamefont {Park}\ \emph {et~al.}(2007)\citenamefont {Park},
  \citenamefont {{Won Lee}}, \citenamefont {Yang}, \citenamefont {Jo},\ and\
  \citenamefont {Moon}}]{Park2007}%
  \BibitemOpen
  \bibfield  {author} {\bibinfo {author} {\bibfnamefont {J.~B.}\ \bibnamefont
  {Park}}, \bibinfo {author} {\bibfnamefont {J.}~\bibnamefont {{Won Lee}}},
  \bibinfo {author} {\bibfnamefont {J.-S.}\ \bibnamefont {Yang}}, \bibinfo
  {author} {\bibfnamefont {H.-H.}\ \bibnamefont {Jo}}, \ and\ \bibinfo {author}
  {\bibfnamefont {H.-T.}\ \bibnamefont {Moon}},\ }\href {\doibase
  10.1016/j.physa.2006.12.042} {\bibfield  {journal} {\bibinfo  {journal}
  {Physica A: Statistical Mechanics and its Applications}\ }\textbf {\bibinfo
  {volume} {379}},\ \bibinfo {pages} {179} (\bibinfo {year} {2007})},\ \Eprint
  {http://arxiv.org/abs/0607283} {arXiv:0607283 [physics]} \BibitemShut
  {NoStop}%
\bibitem [{\citenamefont {Yang}\ \emph {et~al.}(2008)\citenamefont {Yang},
  \citenamefont {Kwak}, \citenamefont {Kaizoji},\ and\ \citenamefont
  {Kim}}]{Yang2008}%
  \BibitemOpen
  \bibfield  {author} {\bibinfo {author} {\bibfnamefont {J.-S.}\ \bibnamefont
  {Yang}}, \bibinfo {author} {\bibfnamefont {W.}~\bibnamefont {Kwak}}, \bibinfo
  {author} {\bibfnamefont {T.}~\bibnamefont {Kaizoji}}, \ and\ \bibinfo
  {author} {\bibfnamefont {I.-m.}\ \bibnamefont {Kim}},\ }\href {\doibase
  10.1140/epjb/e2008-00050-0} {\bibfield  {journal} {\bibinfo  {journal} {The
  European Physical Journal B}\ }\textbf {\bibinfo {volume} {61}},\ \bibinfo
  {pages} {241} (\bibinfo {year} {2008})},\ \Eprint
  {http://arxiv.org/abs/0701179} {arXiv:0701179 [physics]} \BibitemShut
  {NoStop}%
\bibitem [{\citenamefont {Varn}\ \emph {et~al.}(2002)\citenamefont {Varn},
  \citenamefont {Canright},\ and\ \citenamefont {Crutchfield}}]{Varn2002}%
  \BibitemOpen
  \bibfield  {author} {\bibinfo {author} {\bibfnamefont {D.~P.}\ \bibnamefont
  {Varn}}, \bibinfo {author} {\bibfnamefont {G.~S.}\ \bibnamefont {Canright}},
  \ and\ \bibinfo {author} {\bibfnamefont {J.~P.}\ \bibnamefont
  {Crutchfield}},\ }\href {\doibase 10.1103/PhysRevB.66.174110} {\bibfield
  {journal} {\bibinfo  {journal} {Physical Review B}\ }\textbf {\bibinfo
  {volume} {66}},\ \bibinfo {pages} {174110} (\bibinfo {year} {2002})},\
  \Eprint {http://arxiv.org/abs/0203290} {arXiv:0203290 [cond-mat]}
  \BibitemShut {NoStop}%
\bibitem [{\citenamefont {Varn}\ \emph {et~al.}(2013)\citenamefont {Varn},
  \citenamefont {Canright},\ and\ \citenamefont {Crutchfield}}]{Varn2013}%
  \BibitemOpen
  \bibfield  {author} {\bibinfo {author} {\bibfnamefont {D.~P.}\ \bibnamefont
  {Varn}}, \bibinfo {author} {\bibfnamefont {G.~S.}\ \bibnamefont {Canright}},
  \ and\ \bibinfo {author} {\bibfnamefont {J.~P.}\ \bibnamefont
  {Crutchfield}},\ }\href {\doibase 10.1107/S0108767312046582} {\bibfield
  {journal} {\bibinfo  {journal} {Acta Crystallographica Section A}\ }\textbf
  {\bibinfo {volume} {69}},\ \bibinfo {pages} {197} (\bibinfo {year}
  {2013})}\BibitemShut {NoStop}%
\bibitem [{\citenamefont {Varn}\ and\ \citenamefont
  {Crutchfield}(2015)}]{Varn2015}%
  \BibitemOpen
  \bibfield  {author} {\bibinfo {author} {\bibfnamefont {D.~P.}\ \bibnamefont
  {Varn}}\ and\ \bibinfo {author} {\bibfnamefont {J.~P.}\ \bibnamefont
  {Crutchfield}},\ }\href {\doibase 10.1016/j.coche.2014.11.002} {\bibfield
  {journal} {\bibinfo  {journal} {Current Opinion in Chemical Engineering}\
  }\textbf {\bibinfo {volume} {7}},\ \bibinfo {pages} {47} (\bibinfo {year}
  {2015})},\ \Eprint {http://arxiv.org/abs/1409.5930} {arXiv:1409.5930}
  \BibitemShut {NoStop}%
\bibitem [{\citenamefont {Crutchfield}\ and\ \citenamefont
  {Feldman}(1997)}]{Crutchfield1997}%
  \BibitemOpen
  \bibfield  {author} {\bibinfo {author} {\bibfnamefont {J.~P.}\ \bibnamefont
  {Crutchfield}}\ and\ \bibinfo {author} {\bibfnamefont {D.~P.}\ \bibnamefont
  {Feldman}},\ }\href {\doibase 10.1103/PhysRevE.55.R1239} {\bibfield
  {journal} {\bibinfo  {journal} {Physical Review E}\ }\textbf {\bibinfo
  {volume} {55}},\ \bibinfo {pages} {R1239} (\bibinfo {year} {1997})},\ \Eprint
  {http://arxiv.org/abs/9702191} {arXiv:9702191 [cond-mat]} \BibitemShut
  {NoStop}%
\bibitem [{\citenamefont {Suen}\ \emph {et~al.}(2017)\citenamefont {Suen},
  \citenamefont {Thompson}, \citenamefont {Garner}, \citenamefont {Vedral},\
  and\ \citenamefont {Gu}}]{Suen2017}%
  \BibitemOpen
  \bibfield  {author} {\bibinfo {author} {\bibfnamefont {W.~Y.}\ \bibnamefont
  {Suen}}, \bibinfo {author} {\bibfnamefont {J.}~\bibnamefont {Thompson}},
  \bibinfo {author} {\bibfnamefont {A.~J.~P.}\ \bibnamefont {Garner}}, \bibinfo
  {author} {\bibfnamefont {V.}~\bibnamefont {Vedral}}, \ and\ \bibinfo {author}
  {\bibfnamefont {M.}~\bibnamefont {Gu}},\ }\href
  {https://doi.org/10.22331/q-2017-08-11-25} {\bibfield  {journal} {\bibinfo
  {journal} {Quantum}\ }\textbf {\bibinfo {volume} {1}},\ \bibinfo {pages} {25}
  (\bibinfo {year} {2017})},\ \Eprint {http://arxiv.org/abs/1511.05738}
  {arXiv:1511.05738} \BibitemShut {NoStop}%
\bibitem [{\citenamefont {Aghamohammadi}\ \emph
  {et~al.}(2017{\natexlab{a}})\citenamefont {Aghamohammadi}, \citenamefont
  {Mahoney},\ and\ \citenamefont {Crutchfield}}]{Aghamohammadi2017}%
  \BibitemOpen
  \bibfield  {author} {\bibinfo {author} {\bibfnamefont {C.}~\bibnamefont
  {Aghamohammadi}}, \bibinfo {author} {\bibfnamefont {J.~R.}\ \bibnamefont
  {Mahoney}}, \ and\ \bibinfo {author} {\bibfnamefont {J.~P.}\ \bibnamefont
  {Crutchfield}},\ }\href {\doibase 10.1016/j.physleta.2016.12.036} {\bibfield
  {journal} {\bibinfo  {journal} {Physics Letters A}\ }\textbf {\bibinfo
  {volume} {381}},\ \bibinfo {pages} {1223} (\bibinfo {year}
  {2017}{\natexlab{a}})},\ \Eprint {http://arxiv.org/abs/1602.08646}
  {arXiv:1602.08646} \BibitemShut {NoStop}%
\bibitem [{\citenamefont {Aghamohammadi}\ \emph
  {et~al.}(2017{\natexlab{b}})\citenamefont {Aghamohammadi}, \citenamefont
  {Mahoney},\ and\ \citenamefont {Crutchfield}}]{Aghamohammadi2017a}%
  \BibitemOpen
  \bibfield  {author} {\bibinfo {author} {\bibfnamefont {C.}~\bibnamefont
  {Aghamohammadi}}, \bibinfo {author} {\bibfnamefont {J.~R.}\ \bibnamefont
  {Mahoney}}, \ and\ \bibinfo {author} {\bibfnamefont {J.~P.}\ \bibnamefont
  {Crutchfield}},\ }\href {\doibase 10.1038/s41598-017-04928-7} {\bibfield
  {journal} {\bibinfo  {journal} {Scientific Reports}\ }\textbf {\bibinfo
  {volume} {7}},\ \bibinfo {pages} {6735} (\bibinfo {year}
  {2017}{\natexlab{b}})},\ \Eprint {http://arxiv.org/abs/1609.03650}
  {arXiv:1609.03650} \BibitemShut {NoStop}%
\bibitem [{\citenamefont {Jouneghani}\ \emph {et~al.}(2017)\citenamefont
  {Jouneghani}, \citenamefont {Gu}, \citenamefont {Ho}, \citenamefont
  {Thompson}, \citenamefont {Suen}, \citenamefont {Wiseman},\ and\
  \citenamefont {Pryde}}]{Jouneghani2017}%
  \BibitemOpen
  \bibfield  {author} {\bibinfo {author} {\bibfnamefont {F.~G.}\ \bibnamefont
  {Jouneghani}}, \bibinfo {author} {\bibfnamefont {M.}~\bibnamefont {Gu}},
  \bibinfo {author} {\bibfnamefont {J.}~\bibnamefont {Ho}}, \bibinfo {author}
  {\bibfnamefont {J.}~\bibnamefont {Thompson}}, \bibinfo {author}
  {\bibfnamefont {W.~Y.}\ \bibnamefont {Suen}}, \bibinfo {author}
  {\bibfnamefont {H.~M.}\ \bibnamefont {Wiseman}}, \ and\ \bibinfo {author}
  {\bibfnamefont {G.~J.}\ \bibnamefont {Pryde}},\ }\href@noop {} {\bibfield
  {journal} {\bibinfo  {journal} {arXiv preprint}\ } (\bibinfo {year}
  {2017})},\ \Eprint {http://arxiv.org/abs/1711.03661} {arXiv:1711.03661}
  \BibitemShut {NoStop}%
\bibitem [{\citenamefont {Gu}\ \emph {et~al.}(2012)\citenamefont {Gu},
  \citenamefont {Wiesner}, \citenamefont {Rieper},\ and\ \citenamefont
  {Vedral}}]{Gu2012}%
  \BibitemOpen
  \bibfield  {author} {\bibinfo {author} {\bibfnamefont {M.}~\bibnamefont
  {Gu}}, \bibinfo {author} {\bibfnamefont {K.}~\bibnamefont {Wiesner}},
  \bibinfo {author} {\bibfnamefont {E.}~\bibnamefont {Rieper}}, \ and\ \bibinfo
  {author} {\bibfnamefont {V.}~\bibnamefont {Vedral}},\ }\href {\doibase
  10.1038/ncomms1761} {\bibfield  {journal} {\bibinfo  {journal} {Nature
  Communications}\ }\textbf {\bibinfo {volume} {3}},\ \bibinfo {pages} {762}
  (\bibinfo {year} {2012})},\ \Eprint {http://arxiv.org/abs/1102.1994}
  {arXiv:1102.1994} \BibitemShut {NoStop}%
\bibitem [{\citenamefont {Mahoney}\ \emph {et~al.}(2016)\citenamefont
  {Mahoney}, \citenamefont {Aghamohammadi},\ and\ \citenamefont
  {Crutchfield}}]{Mahoney2016}%
  \BibitemOpen
  \bibfield  {author} {\bibinfo {author} {\bibfnamefont {J.~R.}\ \bibnamefont
  {Mahoney}}, \bibinfo {author} {\bibfnamefont {C.}~\bibnamefont
  {Aghamohammadi}}, \ and\ \bibinfo {author} {\bibfnamefont {J.~P.}\
  \bibnamefont {Crutchfield}},\ }\href {\doibase 10.1038/srep20495} {\bibfield
  {journal} {\bibinfo  {journal} {Scientific Reports}\ }\textbf {\bibinfo
  {volume} {6}},\ \bibinfo {pages} {20495} (\bibinfo {year} {2016})},\ \Eprint
  {http://arxiv.org/abs/1508.02760} {arXiv:1508.02760} \BibitemShut {NoStop}%
\bibitem [{\citenamefont {Riechers}\ \emph {et~al.}(2016)\citenamefont
  {Riechers}, \citenamefont {Mahoney}, \citenamefont {Aghamohammadi},\ and\
  \citenamefont {Crutchfield}}]{Riechers2016}%
  \BibitemOpen
  \bibfield  {author} {\bibinfo {author} {\bibfnamefont {P.~M.}\ \bibnamefont
  {Riechers}}, \bibinfo {author} {\bibfnamefont {J.~R.}\ \bibnamefont
  {Mahoney}}, \bibinfo {author} {\bibfnamefont {C.}~\bibnamefont
  {Aghamohammadi}}, \ and\ \bibinfo {author} {\bibfnamefont {J.~P.}\
  \bibnamefont {Crutchfield}},\ }\href {\doibase 10.1103/PhysRevA.93.052317}
  {\bibfield  {journal} {\bibinfo  {journal} {Physical Review A}\ }\textbf
  {\bibinfo {volume} {93}},\ \bibinfo {pages} {052317} (\bibinfo {year}
  {2016})},\ \Eprint {http://arxiv.org/abs/1510.08186} {arXiv:1510.08186}
  \BibitemShut {NoStop}%
\bibitem [{\citenamefont {Aghamohammadi}\ \emph {et~al.}(2018)\citenamefont
  {Aghamohammadi}, \citenamefont {Loomis}, \citenamefont {Mahoney},\ and\
  \citenamefont {Crutchfield}}]{Aghamohammadi2018}%
  \BibitemOpen
  \bibfield  {author} {\bibinfo {author} {\bibfnamefont {C.}~\bibnamefont
  {Aghamohammadi}}, \bibinfo {author} {\bibfnamefont {S.~P.}\ \bibnamefont
  {Loomis}}, \bibinfo {author} {\bibfnamefont {J.~R.}\ \bibnamefont {Mahoney}},
  \ and\ \bibinfo {author} {\bibfnamefont {J.~P.}\ \bibnamefont
  {Crutchfield}},\ }\href {\doibase 10.1103/PhysRevX.8.011025} {\bibfield
  {journal} {\bibinfo  {journal} {Physical Review X}\ }\textbf {\bibinfo
  {volume} {8}},\ \bibinfo {pages} {011025} (\bibinfo {year} {2018})},\ \Eprint
  {http://arxiv.org/abs/1707.09553} {arXiv:1707.09553} \BibitemShut {NoStop}%
\bibitem [{\citenamefont {Gallager}(2013)}]{Gallager2013}%
  \BibitemOpen
  \bibfield  {author} {\bibinfo {author} {\bibfnamefont {R.~G.}\ \bibnamefont
  {Gallager}},\ }\href@noop {} {\emph {\bibinfo {title} {{Stochastic
  Processes}}}}\ (\bibinfo  {publisher} {Cambridge University Press},\ \bibinfo
  {year} {2013})\BibitemShut {NoStop}%
\bibitem [{\citenamefont {Rabiner}\ and\ \citenamefont
  {Juang}(1986)}]{Rabiner1986}%
  \BibitemOpen
  \bibfield  {author} {\bibinfo {author} {\bibfnamefont {L.~R.}\ \bibnamefont
  {Rabiner}}\ and\ \bibinfo {author} {\bibfnamefont {B.~H.}\ \bibnamefont
  {Juang}},\ }\href {\doibase 10.1109/MASSP.1986.1165342} {\bibfield  {journal}
  {\bibinfo  {journal} {IEEE ASSP Magazine}\ }\textbf {\bibinfo {volume} {3}},\
  \bibinfo {pages} {4} (\bibinfo {year} {1986})}\BibitemShut {NoStop}%
\bibitem [{\citenamefont {Shalizi}\ and\ \citenamefont
  {Crutchfield}(2002)}]{Shalizi2002}%
  \BibitemOpen
  \bibfield  {author} {\bibinfo {author} {\bibfnamefont {C.~R.}\ \bibnamefont
  {Shalizi}}\ and\ \bibinfo {author} {\bibfnamefont {J.~P.}\ \bibnamefont
  {Crutchfield}},\ }\href@noop {} {\bibfield  {journal} {\bibinfo  {journal}
  {arXiv preprint}\ ,\ \bibinfo {pages} {cs/0210025}} (\bibinfo {year}
  {2002})},\ \Eprint {http://arxiv.org/abs/0210025} {arXiv:0210025 [cs]}
  \BibitemShut {NoStop}%
\bibitem [{\citenamefont {Strelioff}\ and\ \citenamefont
  {Crutchfield}(2014)}]{Strelioff2014}%
  \BibitemOpen
  \bibfield  {author} {\bibinfo {author} {\bibfnamefont {C.~C.}\ \bibnamefont
  {Strelioff}}\ and\ \bibinfo {author} {\bibfnamefont {J.~P.}\ \bibnamefont
  {Crutchfield}},\ }\href {\doibase 10.1103/PhysRevE.89.042119} {\bibfield
  {journal} {\bibinfo  {journal} {Physical Review E}\ }\textbf {\bibinfo
  {volume} {89}},\ \bibinfo {pages} {042119} (\bibinfo {year} {2014})},\
  \Eprint {http://arxiv.org/abs/1309.1392} {arXiv:1309.1392} \BibitemShut
  {NoStop}%
\bibitem [{\citenamefont {Grassberger}(1986)}]{Grassberger1986}%
  \BibitemOpen
  \bibfield  {author} {\bibinfo {author} {\bibfnamefont {P.}~\bibnamefont
  {Grassberger}},\ }\href {\doibase 10.1007/BF00668821} {\bibfield  {journal}
  {\bibinfo  {journal} {International Journal of Theoretical Physics}\ }\textbf
  {\bibinfo {volume} {25}},\ \bibinfo {pages} {907} (\bibinfo {year}
  {1986})}\BibitemShut {NoStop}%
\bibitem [{\citenamefont {Jozsa}\ and\ \citenamefont
  {Schlienz}(2000)}]{Jozsa2000}%
  \BibitemOpen
  \bibfield  {author} {\bibinfo {author} {\bibfnamefont {R.}~\bibnamefont
  {Jozsa}}\ and\ \bibinfo {author} {\bibfnamefont {J.}~\bibnamefont
  {Schlienz}},\ }\href {\doibase 10.1103/PhysRevA.62.012301} {\bibfield
  {journal} {\bibinfo  {journal} {Physical Review A}\ }\textbf {\bibinfo
  {volume} {62}},\ \bibinfo {pages} {012301} (\bibinfo {year} {2000})},\
  \Eprint {http://arxiv.org/abs/9911009} {arXiv:9911009 [quant-ph]}
  \BibitemShut {NoStop}%
\bibitem [{\citenamefont {Schumacher}(1995)}]{Schumacher1995}%
  \BibitemOpen
  \bibfield  {author} {\bibinfo {author} {\bibfnamefont {B.}~\bibnamefont
  {Schumacher}},\ }\href {\doibase 10.1103/PhysRevA.51.2738} {\bibfield
  {journal} {\bibinfo  {journal} {Physical Review A}\ }\textbf {\bibinfo
  {volume} {51}},\ \bibinfo {pages} {2738} (\bibinfo {year} {1995})},\ \Eprint
  {http://arxiv.org/abs/quant-ph/9809023} {arXiv:quant-ph/9809023 [quant-ph]}
  \BibitemShut {NoStop}%
\bibitem [{\citenamefont {Elliott}\ and\ \citenamefont
  {Gu}(2018)}]{Elliott2018}%
  \BibitemOpen
  \bibfield  {author} {\bibinfo {author} {\bibfnamefont {T.~J.}\ \bibnamefont
  {Elliott}}\ and\ \bibinfo {author} {\bibfnamefont {M.}~\bibnamefont {Gu}},\
  }\href {\doibase 10.1038/s41534-018-0064-4} {\bibfield  {journal} {\bibinfo
  {journal} {npj Quantum Information}\ }\textbf {\bibinfo {volume} {4}},\
  \bibinfo {pages} {18} (\bibinfo {year} {2018})},\ \Eprint
  {http://arxiv.org/abs/1704.04231} {arXiv:1704.04231} \BibitemShut {NoStop}%
\bibitem [{\citenamefont {Garner}\ \emph {et~al.}(2017)\citenamefont {Garner},
  \citenamefont {Liu}, \citenamefont {Thompson}, \citenamefont {Vedral},\ and\
  \citenamefont {Gu}}]{Garner2017}%
  \BibitemOpen
  \bibfield  {author} {\bibinfo {author} {\bibfnamefont {A.~J.~P.}\
  \bibnamefont {Garner}}, \bibinfo {author} {\bibfnamefont {Q.}~\bibnamefont
  {Liu}}, \bibinfo {author} {\bibfnamefont {J.}~\bibnamefont {Thompson}},
  \bibinfo {author} {\bibfnamefont {V.}~\bibnamefont {Vedral}}, \ and\ \bibinfo
  {author} {\bibfnamefont {M.}~\bibnamefont {Gu}},\ }\href {\doibase
  10.1088/1367-2630/aa82df} {\bibfield  {journal} {\bibinfo  {journal} {New
  Journal of Physics}\ }\textbf {\bibinfo {volume} {19}},\ \bibinfo {pages}
  {103009} (\bibinfo {year} {2017})},\ \Eprint
  {http://arxiv.org/abs/1609.04408} {arXiv:1609.04408} \BibitemShut {NoStop}%
\bibitem [{\citenamefont {Thompson}\ \emph
  {et~al.}(2017{\natexlab{a}})\citenamefont {Thompson}, \citenamefont {Garner},
  \citenamefont {Mahoney}, \citenamefont {Crutchfield}, \citenamefont
  {Vedral},\ and\ \citenamefont {Gu}}]{Thompson2017a}%
  \BibitemOpen
  \bibfield  {author} {\bibinfo {author} {\bibfnamefont {J.}~\bibnamefont
  {Thompson}}, \bibinfo {author} {\bibfnamefont {A.~J.~P.}\ \bibnamefont
  {Garner}}, \bibinfo {author} {\bibfnamefont {J.~R.}\ \bibnamefont {Mahoney}},
  \bibinfo {author} {\bibfnamefont {J.~P.}\ \bibnamefont {Crutchfield}},
  \bibinfo {author} {\bibfnamefont {V.}~\bibnamefont {Vedral}}, \ and\ \bibinfo
  {author} {\bibfnamefont {M.}~\bibnamefont {Gu}},\ }\href@noop {} {\bibfield
  {journal} {\bibinfo  {journal} {arXiv preprint}\ ,\ \bibinfo {pages}
  {1712.02368}} (\bibinfo {year} {2017}{\natexlab{a}})},\ \Eprint
  {http://arxiv.org/abs/1712.02368} {arXiv:1712.02368} \BibitemShut {NoStop}%
\bibitem [{\citenamefont {Palsson}\ \emph {et~al.}(2017)\citenamefont
  {Palsson}, \citenamefont {Gu}, \citenamefont {Ho}, \citenamefont {Wiseman},\
  and\ \citenamefont {Pryde}}]{Palsson2017}%
  \BibitemOpen
  \bibfield  {author} {\bibinfo {author} {\bibfnamefont {M.~S.}\ \bibnamefont
  {Palsson}}, \bibinfo {author} {\bibfnamefont {M.}~\bibnamefont {Gu}},
  \bibinfo {author} {\bibfnamefont {J.}~\bibnamefont {Ho}}, \bibinfo {author}
  {\bibfnamefont {H.~M.}\ \bibnamefont {Wiseman}}, \ and\ \bibinfo {author}
  {\bibfnamefont {G.~J.}\ \bibnamefont {Pryde}},\ }\href {\doibase
  10.1126/sciadv.1601302} {\bibfield  {journal} {\bibinfo  {journal} {Science
  Advances}\ }\textbf {\bibinfo {volume} {3}},\ \bibinfo {pages} {e1601302}
  (\bibinfo {year} {2017})},\ \Eprint {http://arxiv.org/abs/1602.05683}
  {arXiv:1602.05683} \BibitemShut {NoStop}%
\bibitem [{\citenamefont {Marzen}\ and\ \citenamefont
  {Crutchfield}(2017)}]{Marzen2016}%
  \BibitemOpen
  \bibfield  {author} {\bibinfo {author} {\bibfnamefont {S.}~\bibnamefont
  {Marzen}}\ and\ \bibinfo {author} {\bibfnamefont {J.~P.}\ \bibnamefont
  {Crutchfield}},\ }\href
  {https://link.springer.com/article/10.1007{\%}2Fs10955-017-1793-z
  http://arxiv.org/abs/1611.01099} {\bibfield  {journal} {\bibinfo  {journal}
  {Journal of Statistical Physics}\ }\textbf {\bibinfo {volume} {168}},\
  \bibinfo {pages} {109} (\bibinfo {year} {2017})},\ \Eprint
  {http://arxiv.org/abs/1611.01099} {arXiv:1611.01099} \BibitemShut {NoStop}%
\bibitem [{\citenamefont {Barnett}\ and\ \citenamefont
  {Crutchfield}(2015)}]{Barnett2015}%
  \BibitemOpen
  \bibfield  {author} {\bibinfo {author} {\bibfnamefont {N.}~\bibnamefont
  {Barnett}}\ and\ \bibinfo {author} {\bibfnamefont {J.~P.}\ \bibnamefont
  {Crutchfield}},\ }\href {\doibase 10.1007/s10955-015-1327-5} {\bibfield
  {journal} {\bibinfo  {journal} {Journal of Statistical Physics}\ }\textbf
  {\bibinfo {volume} {161}},\ \bibinfo {pages} {404} (\bibinfo {year}
  {2015})},\ \Eprint {http://arxiv.org/abs/1412.2690} {arXiv:1412.2690}
  \BibitemShut {NoStop}%
\bibitem [{\citenamefont {Thompson}\ \emph
  {et~al.}(2017{\natexlab{b}})\citenamefont {Thompson}, \citenamefont {Garner},
  \citenamefont {Vedral},\ and\ \citenamefont {Gu}}]{Thompson2017}%
  \BibitemOpen
  \bibfield  {author} {\bibinfo {author} {\bibfnamefont {J.}~\bibnamefont
  {Thompson}}, \bibinfo {author} {\bibfnamefont {A.~J.~P.}\ \bibnamefont
  {Garner}}, \bibinfo {author} {\bibfnamefont {V.}~\bibnamefont {Vedral}}, \
  and\ \bibinfo {author} {\bibfnamefont {M.}~\bibnamefont {Gu}},\ }\href
  {\doibase 10.1038/s41534-016-0001-3} {\bibfield  {journal} {\bibinfo
  {journal} {npj Quantum Information}\ }\textbf {\bibinfo {volume} {3}}
  (\bibinfo {year} {2017}{\natexlab{b}}),\ 10.1038/s41534-016-0001-3},\ \Eprint
  {http://arxiv.org/abs/1601.05420} {arXiv:1601.05420} \BibitemShut {NoStop}%
\bibitem [{\citenamefont {Yang}\ \emph {et~al.}(2018)\citenamefont {Yang},
  \citenamefont {Binder}, \citenamefont {Narasimhachar},\ and\ \citenamefont
  {Gu}}]{Yang2018}%
  \BibitemOpen
  \bibfield  {author} {\bibinfo {author} {\bibfnamefont {C.}~\bibnamefont
  {Yang}}, \bibinfo {author} {\bibfnamefont {F.~C.}\ \bibnamefont {Binder}},
  \bibinfo {author} {\bibfnamefont {V.}~\bibnamefont {Narasimhachar}}, \ and\
  \bibinfo {author} {\bibfnamefont {M.}~\bibnamefont {Gu}},\ }\href@noop {}
  {\bibfield  {journal} {\bibinfo  {journal} {arXiv preprint}\ ,\ \bibinfo
  {pages} {1803.08220}} (\bibinfo {year} {2018})},\ \Eprint
  {http://arxiv.org/abs/1803.08220} {arXiv:1803.08220} \BibitemShut {NoStop}%
\bibitem [{\citenamefont {Gerschgorin}(1931)}]{Gerschgorin1931}%
  \BibitemOpen
  \bibfield  {author} {\bibinfo {author} {\bibfnamefont {S.}~\bibnamefont
  {Gerschgorin}},\ }\href@noop {} {\bibfield  {journal} {\bibinfo  {journal}
  {Bulletin de l'Acad{\'{e}}mie des Sciences de l'URSS}\ }\textbf {\bibinfo
  {volume} {1931}},\ \bibinfo {pages} {749} (\bibinfo {year}
  {1931})}\BibitemShut {NoStop}%
\end{thebibliography}
\end{document}